\documentclass[aip, pof, reprint]{revtex4-2}
\usepackage{amsfonts,amsmath,bm,amssymb,color}
\usepackage{graphicx}
\usepackage[colorlinks=true, citecolor=blue, linkcolor=blue]{hyperref}

\usepackage{subcaption} 

\begin{document}

\title{The effects of no-slip boundaries and external force torque on two-dimensional turbulence in a square domain}

\author{Alisa Shikanian}
\affiliation{Landau Institute for Theoretical Physics, Russian Academy of Sciences, 1-A Akademika Semenova av., 142432 Chernogolovka, Russia}
\affiliation{HSE University, Faculty of Physics, Myasnitskaya 20, 101000 Moscow, Russia}

\author{Vladimir Parfenyev}\email{parfenius@gmail.com}
\affiliation{Landau Institute for Theoretical Physics, Russian Academy of Sciences, 1-A Akademika Semenova av., 142432 Chernogolovka, Russia}
\affiliation{HSE University, Faculty of Physics, Myasnitskaya 20, 101000 Moscow, Russia}

\date{\today}

\begin{abstract}
We study two-dimensional turbulence in a square no-slip domain without bottom drag using direct numerical simulations. The dynamics are shown to depend strongly on the torque $M$ of the external forcing. When $M$ is relatively large, a long-lived coherent vortex forms at the domain center, establishing a persistent angular momentum. At lower torques, the angular momentum undergoes random sign reversals due to spontaneous switching of the central vortex circulation, though it predominantly aligns with the torque direction. We investigate the transition between these regimes by smoothly varying $M$, observing that the time-averaged angular momentum of the system follows $\langle \mathcal{L} \rangle \propto M^{1/3}$. A significant part of the energy dissipates near the domain boundaries, requiring a revision of scaling laws for homogeneous systems. New scaling relations are proposed and they enable velocity profiles in the boundary layers to collapse onto a universal curve for simulations with varying fluid viscosities and forcing amplitudes. The velocity profiles feature a linear viscous sublayer followed by a short logarithmic region. The boundary layer thickness $\delta$ scales with the large-scale Reynolds number $Re$ as $\delta \sim L \cdot Re^{-1/2}$, indicating that the friction at no-slip walls is not sufficient to halt the inverse energy cascade before it reaches the system size $L$. The results highlight how no-slip boundaries and forcing asymmetry govern the dynamics of large-scale coherent structures in confined two-dimensional turbulence.

\end{abstract}

\maketitle

\section{Introduction}

The behavior of two-dimensional (2D) turbulent flow differs notably from its three-dimensional (3D) counterpart. This distinction arises because, in the absence of dissipation and forcing, the 2D Navier-Stokes equation has two quadratic integrals of motion: the energy $E = \frac{1}{S}\int dx dy \, \bm v^2/2$ and the enstrophy $Z = \frac{1}{S}\int dx dy \, \omega^2/2$, where $\bm v$ represents the 2D velocity, $\omega = \partial_x v_y - \partial_y v_x$ is the vorticity, and $S$ is the system area. Consequently, hydrodynamic nonlinearity leads to the simultaneous formation of two cascades~\cite{boffetta2012two, boffetta2010evidence}: the flow of enstrophy is directed to small scales (direct cascade) and it stops due to viscosity, and the flow of energy is directed in the opposite direction (inverse cascade) and it is halted by the large-scale dissipation, which often takes the form of linear bottom drag. When the large-scale dissipation is small enough, the inverse energy cascade can propagate to system-size scales, resulting in the formation of coherent structures that were observed already in the first laboratory~\cite{sommeria1986experimental, paret1998intermittency} and numerical~\cite{smith1993bose, smithr1994finite, borue1994inverse} experiments. In contrast, 3D turbulence is characterized solely by a direct energy cascade~\cite{frisch1995turbulence}, which progresses from pumping scale towards smaller scales, and enstrophy is not conserved, meaning its cascade does not exist.

The structure of large-scale coherent flows in 2D turbulence is strongly influenced by the geometry of the domain and the boundary conditions. The most thoroughly studied configurations involve square~\cite{chertkov2007dynamics, chan2012dynamics, laurie2014universal, frishman2018turbulence, parfenyev2022profile, parfenyev2024statistical, vankan2025two} and rectangular~\cite{frishman2017jets, xu2024fluctuation} domains with periodic boundary conditions, primarily due to the simplicity of numerically modeling such systems. In these cases, efficient pseudospectral methods based on the fast Fourier transform can be employed for calculations. Additionally, the application of hyperviscosity reduces the range of the direct enstrophy cascade, facilitating numerical simulations on relatively coarse grids. In particular, direct numerical simulations (DNS) reveal that in a square domain a dipole of coherent vortices forms, exhibiting a universal structure in the asymptotic limit~\cite{laurie2014universal, kolokolov2016structure, frishman2017culmination}, and in rectangular domains with an aspect ratio $\gtrsim 1.1$ unidirectional jets predominate~\cite{xu2024fluctuation}.

No-slip boundaries violate the spatial homogeneity of the system, leading to the formation of thin boundary layers near the edges, which can notably influence the energy and enstrophy balances~\cite{clercx2009}. Unlike flows in double periodic domains, large-scale coherent structures adopt a distinct configuration. Both laboratory~\cite{sommeria1986experimental, paret1998intermittency, xia2009spectrally, bardoczi2014experimental, orlov2018large, zhu2024flow} and numerical~\cite{molenaar2004angular, doludenko2021coherent} experiments conducted in square cells reveal the development of a large-scale vortex at the center, surrounded by a shielding ring of opposite vorticity, ensuring that the total circulation of the flow remains zero. Remarkably, in some experiments, the vortex at the center persists for extended periods~\cite{xia2009spectrally, bardoczi2014experimental, doludenko2021coherent}, while in others, it undergoes random destruction and reformation, with the direction of its rotation varying arbitrarily. The latter behavior was first observed in laboratory experiments of J. Sommeria~\cite{sommeria1986experimental}, where reversal frequency was found to diminish with decreasing bottom friction $\alpha$, vanishing entirely as $\alpha \to 0$. However, subsequent numerical work~\cite{molenaar2004angular} demonstrated that reversals can also persist in the system without bottom drag, but only when 2D turbulence is excited by stochastic time-dependent forcing. While the exact reversal mechanism is still unclear, a qualitative explanation suggests that it arises when boundary layers separate and twist, forming small-scale strong vortices. These vortices may then propagate toward the cell's center, disrupting the dominant large-scale vortex and triggering its collapse.

Here, we complement recent studies by exploring 2D turbulence excited by a steady forcing in a square no-slip domain without bottom drag using DNS at high spatial resolution up to $8192^2$. We found that the system behavior strongly depends on whether the external force driving the turbulence has nonzero torque. If torque is present, a long-lived coherent vortex forms at the center, in qualitative agreement with Refs.~\onlinecite{xia2009spectrally, bardoczi2014experimental, doludenko2021coherent}. If no torque is applied, we observe that the central vortex is intermittently destroyed and reformed, exhibiting random reversals. This occurs in the absence of bottom drag and under steady forcing, thereby generalizing the findings of Refs.~\onlinecite{sommeria1986experimental, molenaar2004angular}. We tracked the transition between these regimes by varying the external force torque $M$ as a control parameter. We found that when the torque is relatively large, the system maintains a persistent angular momentum and the central vortex remains stable. At lower torque values, the angular momentum undergoes random sign reversals caused by spontaneous changes in the circulation of the central vortex, although it tends to align with the direction of the applied torque. In the absence of torque, the system exhibits no preferred rotation and central vortices of both signs persist with equal likelihood.
Notably, the time-averaged angular momentum of the system follows an approximate dependence $\langle \mathcal{L} \rangle \propto M^{1/3}$, with no evidence of different behavioral regimes.

We also analyzed the energy balance in the system and found that significant part of energy dissipation occurs near the domain boundaries. This distinguishes the dynamics from those in homogeneous turbulence with periodic boundary conditions and necessitates a revision of the scaling relations. We investigated the velocity structure within the boundary layers and found that it exhibits logarithmic behavior with a linear viscous sublayer. These findings are consistent with both experimental measurements of velocity profiles in soap films~\cite{samanta2014scaling} and prior numerical studies employing coarser grids~\cite{doludenko2021coherent}. The revised scaling relations enable the velocity profiles from DNS with varying fluid viscosities and forcing amplitudes to collapse onto a universal master curve. The boundary layer thickness scales as $\delta \sim L \cdot Re^{-1/2}$, where $L$ is the system size and $Re$ is the large-scale Reynolds number. In particular, this means that friction near the boundaries is not strong enough to stop the inverse energy cascade before it reaches the system scale $L$. The obtained results provide new insights into the role of boundary layers and external force torque in shaping the dynamics of 2D turbulent flows.

\begin{table}[t]
\caption{\label{tab:1}Parameters for the DNS runs: grid resolution, statistics collection time $T$, data saving interval $\Delta t$, fluid viscosity $\nu$, forcing amplitude $f_0$, torque of external force $M$, mean energy $\langle E \rangle$, pumping power $\varepsilon$, energy dissipation rate $2 \nu \langle Z \rangle$, where $\langle Z \rangle$ is the mean enstrophy, and large-scale Reynolds number $Re$.}
\begin{ruledtabular}
\begin{tabular}{ccccccccccc}
         run & grid & $T$ & $\Delta t$ & $\nu$ & $f_0$ & $M$ & $\langle E \rangle$ & $\varepsilon$ & $2 \nu \langle Z \rangle$ & $Re=UL/\nu$ \\ \hline
$A$ & $4096$ & $2000$ & $1.5$ & $2 \cdot 10^{-4}$ & $0.1$ & $\mathbf{4.0 \cdot 10^{-2}}$ & $0.97$ & $5.66 \cdot 10^{-3}$ & $5.61 \cdot 10^{-3}$ & $4.4 \cdot 10^4$ \\
$B$ & $4096$ & $2000$ & $1.5$ & $2 \cdot 10^{-4}$ & $0.1$ & $\mathbf{0}$ & $0.56$ & $4.59 \cdot 10^{-3}$ & $4.58 \cdot 10^{-3}$ & $3.3 \cdot 10^4$ \\ \hline
$V_0 F_2$ & $2048$ & $4000$ & $2$ & $\mathbf{8 \cdot 10^{-4}}$ & $0.1$ & $0$ & $0.22$ & $5.69 \cdot 10^{-3}$ & $5.69 \cdot 10^{-3}$ & $0.5 \cdot 10^4$ \\
$V_1 F_2$ & $4096$ & $4000$ & $2$ & $\mathbf{4 \cdot 10^{-4}}$ & $0.1$ & $0$ & $0.35$ & $4.78 \cdot 10^{-3}$ & $4.71 \cdot 10^{-3}$ & $1.3 \cdot 10^4$ \\
$V_2 F_2$ & $4096$ & $4000$ & $2$ & $\mathbf{2 \cdot 10^{-4}}$ & $0.1$ & $0$ & $0.53$ & $4.62 \cdot 10^{-3}$ & $4.62 \cdot 10^{-3}$ & $3.2 \cdot 10^4$ \\
$V_3 F_2$ & $4096$ & $5000$ & $2$ & $\mathbf{1 \cdot 10^{-4}}$ & $0.1$ & $0$ & $0.74$ & $4.62 \cdot 10^{-3}$ & $4.61 \cdot 10^{-3}$ & $7.7 \cdot 10^4$ \\
$V_4 F_2$ & $8192$ & $1800$ & $1$ & $\mathbf{0.5 \cdot 10^{-4}}$ & $0.1$ & $0$ & $0.91$ & $4.43 \cdot 10^{-3}$ & $4.24 \cdot 10^{-3}$ & $17.0 \cdot 10^4$ \\
$V_5 F_2$ & $8192$ & $1800$ & $1$ & $\mathbf{0.25 \cdot 10^{-4}}$ & $0.1$ & $0$ & $1.59$ & $3.69 \cdot 10^{-3}$ & $3.35 \cdot 10^{-3}$ & $44.8 \cdot 10^4$ \\ \hline
$V_2 F_0$ & $2048$ & $4000$ & $2$ & $2 \cdot 10^{-4}$ & $\mathbf{0.025}$ & $0$ & $0.09$ & $0.61 \cdot 10^{-3}$ & $0.61 \cdot 10^{-3}$ & $1.3 \cdot 10^4$ \\
$V_2 F_1$ & $4096$ & $4000$ & $2$ & $2 \cdot 10^{-4}$ & $\mathbf{0.05}$ & $0$ & $0.20$ & $1.56 \cdot 10^{-3}$ & $1.57 \cdot 10^{-3}$ & $2.0 \cdot 10^4$ \\
$V_2 F_2$ & $4096$ & $4000$ & $2$ & $2 \cdot 10^{-4}$ & $\mathbf{0.1}$ & $0$ & $0.53$ & $4.62 \cdot 10^{-3}$ & $4.62 \cdot 10^{-3}$ & $3.2 \cdot 10^4$ \\
$V_2 F_3$ & $4096$ & $4000$ & $2$ & $2 \cdot 10^{-4}$ & $\mathbf{0.2}$ & $0$ & $1.21$ & $13.00 \cdot 10^{-3}$ & $13.08 \cdot 10^{-3}$ & $4.9 \cdot 10^4$ \\ \hline
$X_0$ & $1024$ & $10^5$ & $20$ & $2 \cdot 10^{-4}$ & $0.1$ & $\mathbf{0}$ & $0.74$ & $4.34 \cdot 10^{-3}$ & $4.24 \cdot 10^{-3}$ & $3.8 \cdot 10^4$ \\
$X_1$ & $1024$ & $10^5$ & $20$ & $2 \cdot 10^{-4}$ & $0.1$ & $\mathbf{0.07 \cdot 10^{-2}}$ & $0.76$ & $4.30 \cdot 10^{-3}$ & $4.29 \cdot 10^{-3}$ & $3.9 \cdot 10^4$ \\
$X_2$ & $1024$ & $10^5$ & $20$ & $2 \cdot 10^{-4}$ & $0.1$ & $\mathbf{0.15 \cdot 10^{-2}}$ & $0.76$ & $4.25 \cdot 10^{-3}$ & $4.31 \cdot 10^{-3}$ & $3.9 \cdot 10^4$ \\
$X_3$ & $1024$ & $10^5$ & $20$ & $2 \cdot 10^{-4}$ & $0.1$ & $\mathbf{0.27 \cdot 10^{-2}}$ & $0.75$ & $4.35 \cdot 10^{-3}$ & $4.24 \cdot 10^{-3}$ & $3.8 \cdot 10^4$ \\
$X_4$ & $1024$ & $10^5$ & $20$ & $2 \cdot 10^{-4}$ & $0.1$ & $\mathbf{0.59 \cdot 10^{-2}}$ & $0.76$ & $4.25 \cdot 10^{-3}$ & $4.27 \cdot 10^{-3}$ & $3.9 \cdot 10^4$ \\
$X_5$ & $1024$ & $10^5$ & $20$ & $2 \cdot 10^{-4}$ & $0.1$ & $\mathbf{1.00 \cdot 10^{-2}}$ & $0.75$ & $4.41 \cdot 10^{-3}$ & $4.42 \cdot 10^{-3}$ & $3.9 \cdot 10^4$ \\
$X_6$ & $1024$ & $10^5$ & $20$ & $2 \cdot 10^{-4}$ & $0.1$ & $\mathbf{1.48 \cdot 10^{-2}}$ & $0.77$ & $4.52 \cdot 10^{-3}$ & $4.62 \cdot 10^{-3}$ & $3.9 \cdot 10^4$ \\
$X_7$ & $1024$ & $10^5$ & $20$ & $2 \cdot 10^{-4}$ & $0.1$ & $\mathbf{2.00 \cdot 10^{-2}}$ & $0.79$ & $4.80 \cdot 10^{-3}$ & $4.82 \cdot 10^{-3}$ & $3.9 \cdot 10^4$ \\
$X_8$ & $1024$ & $10^5$ & $20$ & $2 \cdot 10^{-4}$ & $0.1$ & $\mathbf{2.52 \cdot 10^{-2}}$ & $0.84$ & $5.10 \cdot 10^{-3}$ & $5.06 \cdot 10^{-3}$ & $4.1 \cdot 10^4$ \\
$X_9$ & $1024$ & $10^5$ & $20$ & $2 \cdot 10^{-4}$ & $0.1$ & $\mathbf{3.00 \cdot 10^{-2}}$ & $0.88$ & $5.28 \cdot 10^{-3}$ & $5.25 \cdot 10^{-3}$ & $4.2 \cdot 10^4$ \\
$X_{10}$ & $1024$ & $10^5$ & $20$ & $2 \cdot 10^{-4}$ & $0.1$ & $\mathbf{3.41 \cdot 10^{-2}}$ & $0.93$ & $5.46 \cdot 10^{-3}$ & $5.51 \cdot 10^{-3}$ & $4.3 \cdot 10^4$ \\
$X_{11}$ & $1024$ & $10^5$ & $20$ & $2 \cdot 10^{-4}$ & $0.1$ & $\mathbf{3.73 \cdot 10^{-2}}$ & $0.96$ & $5.67 \cdot 10^{-3}$ & $5.65 \cdot 10^{-3}$ & $4.4 \cdot 10^4$ \\
$X_{12}$ & $1024$ & $10^5$ & $20$ & $2 \cdot 10^{-4}$ & $0.1$ & $\mathbf{4.00 \cdot 10^{-2}}$ & $0.99$ & $5.86 \cdot 10^{-3}$ & $5.86 \cdot 10^{-3}$ & $4.4 \cdot 10^4$ \\
\end{tabular}
\end{ruledtabular}
\end{table}

\section{Numerical Methods}

We solve the incompressible forced 2D Navier-Stokes equation for a fluid with unit density
\begin{equation}\label{eq:NS-equation}
\partial_t \bm v + (\bm v \cdot \nabla) \bm v = - \nabla p + \nu \nabla^2 \bm v + \bm f, \quad \nabla \cdot \bm v = 0,
\end{equation}
where $\bm v$ is 2D velocity, $p$ is the pressure, and $\nu$ is the kinematic viscosity. The domain is a no-slip square box of size $L=2 \pi$. The external force $\bm f$ is stationary in time and periodic in space; it contains harmonics with wave number $k_f=5$. This type of pumping is motivated by laboratory experiments on the study of turbulent flows in thin electrolyte layers~\cite{sommeria1986experimental, paret1998intermittency, xia2009spectrally, bardoczi2014experimental, orlov2018large, zhu2024flow}. However, in contrast to these studies, we consider a system without bottom drag in order to isolate and examine the effects of viscous friction at no-slip walls. The absence of bottom drag is relevant, e.g., for soap films under low air pressure or flows on superhydrophobic surfaces~\cite{rothstein2010slip}.

DNS results are obtained by integrating (\ref{eq:NS-equation}) using the Oceananigans.jl code~\cite{wagner2025high}. It solves these equations on GPU adopting a finite volume discretization, and we use a centered 2nd-order advection scheme and a 3rd-order Runge-Kutta time-stepping method. The integration time steps are chosen adaptively with a Courant number of $0.5-0.7$, which satisfies the convergence condition of the numerical scheme. The grid is uniform in space and has resolution up to $8192^2$. The grid size should be sufficiently small to resolve the boundary layers at no-slip edges and the scale of enstrophy dissipation in the direct cascade. Detailed information on the numerical implementation and algorithms can be found in Ref.~\onlinecite{wagner2025high} and in the Oceananigans.jl documentation.

In all simulations, the initial condition is a state of rest and each simulation is run until the system reaches a non-equilibrium stationary state, observed by the saturation of the total kinetic energy. After this, statistics are collected for some time $T$, and we output data at every $\Delta t$. In this work, we are mostly interested in the properties of turbulent flow in a statistically stationary state, so the transient regime is not analyzed in detail. The parameters of DNS runs are summarized in Table~\ref{tab:1}. The high-resolution DNS runs $A$ and $B$ are designed to compare system dynamics with and without external force torque, as detailed in Sections~\ref{eq:sec3a}-\ref{eq:sec3b}. Subsequent DNS runs $V_{\#} F_{\#}$ test the scaling relations derived in Section~\ref{eq:sec3c}, where $V$ denotes viscosity variations and $F$ represents forcing amplitude adjustments. Lastly, the $X_{\#}$ series explores gradual variations in external torque in Section~\ref{eq:sec3d}. These latter simulations are distinguished by their extended duration, enabling observation of multiple reversals in the central vortex’s rotation direction. To accommodate their long timescales, these runs employ coarser spatial resolution.

\section{Results}

\subsection{The role of external force torque}\label{eq:sec3a}

We begin the presentation of DNS results by examining two cases, denoted as runs $A$ and $B$, which differ in terms of the external forces applied. In the first case, the external force is given by
\begin{equation}
\bm f_A = [f_0 \sin (k_f y), -f_0 \sin(k_f x)],
\end{equation}
which generates a non-zero torque relative to the cell center $\bm r_O$. Specifically, the torque is $M = \frac{1}{L^2}\int d^2 \bm r\, [(\bm r - \bm r_O)\times \bm f_A]_z = 2f_0/k_f$. In the second case, the external force is
\begin{equation}\label{eq:fb}
\bm f_B = [f_0 \cos (k_f y), -f_0 \cos(k_f x)],
\end{equation}
and it results in a zero torque. The forcing parameters are set to $f_0 = 0.1$ and $k_f = 5$, the fluid viscosity is $\nu = 2 \cdot 10^{-4}$, and the simulations are performed on a $4096^2$ grid.

\begin{figure}[b]
    \centering
    \includegraphics[width=0.6\linewidth]{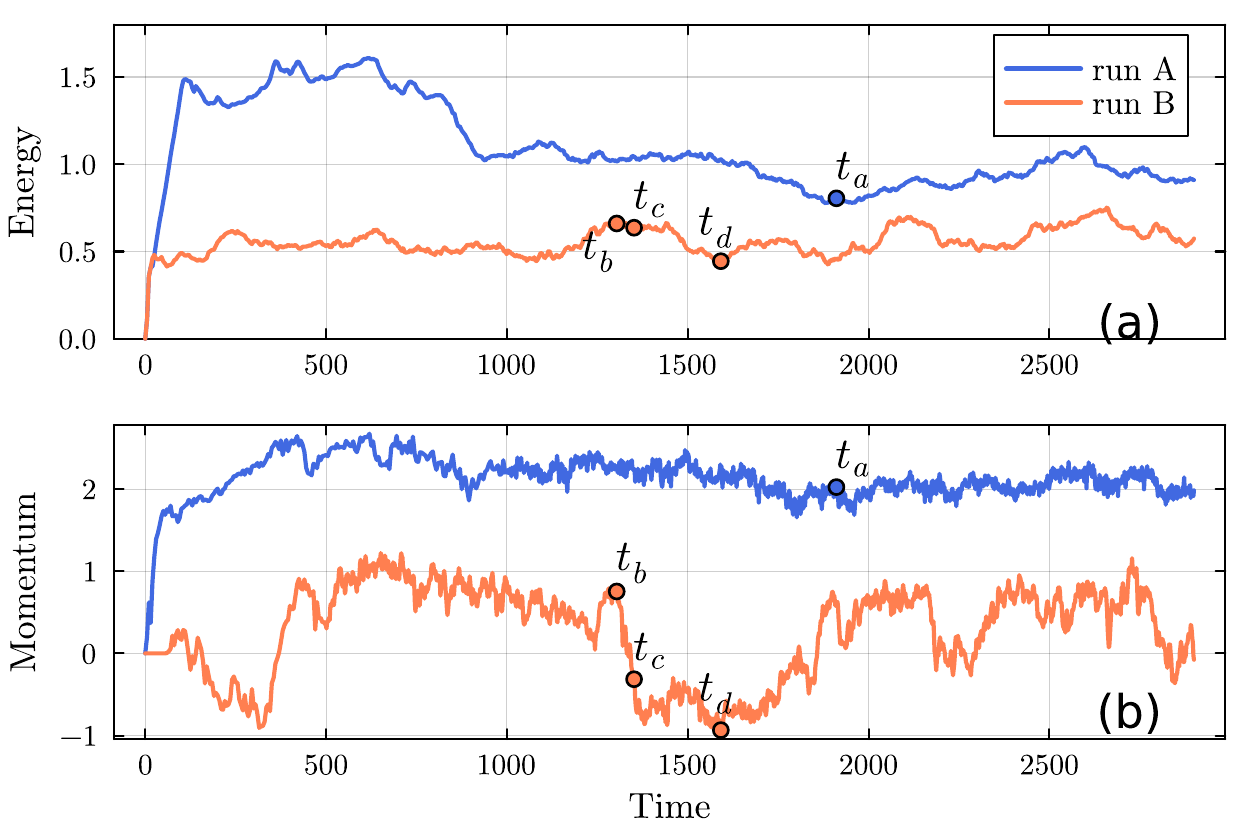}
    \caption{The dependence of the kinetic energy (a) and the angular momentum (b) of the system on time for DNS runs $A$ and $B$.}
    \label{fig:1}
\end{figure}

Figure~\ref{fig:1}a shows the dependence of the kinetic energy of the system $E = \frac{1}{L^2}\int dxdy\, \bm v^2/2$ on time. When torque is applied, the system reaches a statistical steady-state after approximately $10^3$ time units, passing through a phase of energy reduction. In the steady-state, the energy fluctuates noticeably, and its mean value $\langle E \rangle \approx 0.97$ corresponds to a Reynolds number of $Re = U L/\nu \approx 4.4\cdot 10^4$, where $U = \sqrt{2 \langle E \rangle}$. In contrast, when no torque is applied, the system attains a steady-state more quickly, and there is no phase of energy reduction. The mean value of energy $\langle E \rangle \approx 0.56$ turns out to be lower, corresponding to a Reynolds number of $Re \approx 3.3 \cdot 10^4$. In both cases, the flow remains turbulent.

\begin{figure}[t]
    \centering
    \includegraphics[width=0.9\linewidth]{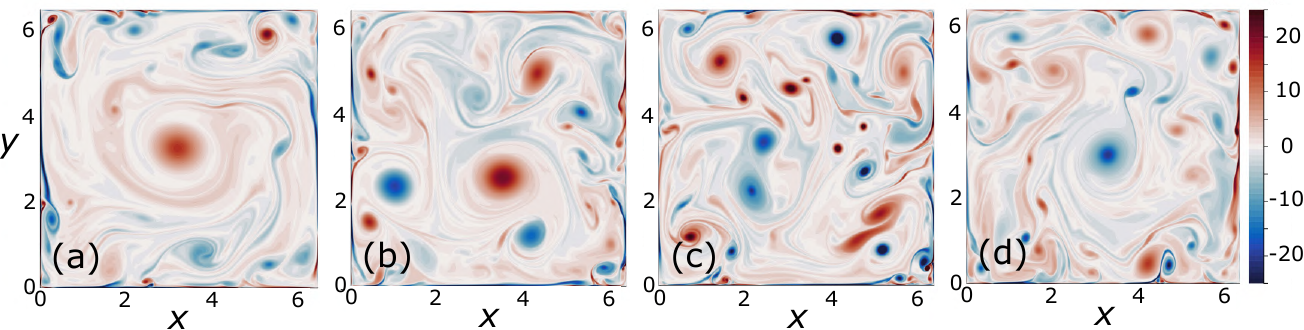}
    \caption{The vorticity distribution at various time instances marked in Fig.~\ref{fig:1}, with panel (a) corresponding to DNS run $A$ and panels (b)-(d) corresponding to DNS run $B$.}
    \label{fig:2}
\end{figure}

The main difference between the two DNS runs is manifested in the dependence of the angular momentum $\mathcal{L} = \frac{1}{L^2}\int d^2 \bm r\, [(\bm r - \bm r_O)\times \bm v]_z$ on time, see Fig.~\ref{fig:1}b. In DNS run $A$, the external force torque leads to a violation of symmetry, and the angular momentum of the system in a steady-state has a well-defined positive value. In this regime, a large-scale coherent vortex is constantly located near the center of the domain, see Fig.~\ref{fig:2}a, consistent with previous observations reported in Ref.~\onlinecite{doludenko2021coherent}. In a more symmetrical case of DNS run $B$, when the external force has no torque, the angular momentum of the system constantly changes its sign, see Fig.~\ref{fig:1}b. Accordingly, a large-scale vortex in the center of the domain disintegrates from time to time, after which it is born again, but the direction of its rotation can be arbitrary. This type of behavior was also observed previously in numerical simulations~\cite{molenaar2004angular}, but with stochastic time-dependent forcing and at lower spatial resolution. Figures~\ref{fig:2}b-\ref{fig:2}d illustrate the distribution of vorticity in the system at different time instances marked in Fig.~\ref{fig:1} for this regime. Note that the angular momentum can be expressed in terms of the vorticity as $\mathcal{L} = -\frac{1}{2L^2} \int d^2 \bm r\, (\bm r - \bm r_O)^2 \omega$. Due to no-slip boundary conditions, the total circulation vanishes $\oint_{\partial D} d \bm s \cdot \bm v = \int d^2 \bm r\, \omega = 0$, which implies that any central large-scale vortex must be accompanied by regions of opposite vorticity. These regions, being farther from the center $\bm r_O$, contribute more heavily to the angular momentum due to the $(\bm r - \bm r_O)^2$ weighting in the integral. Taking into account the general minus sign in front of the integral, one can conclude that the sign of the angular momentum reflects the direction of rotation of the central large-scale vortex.

\subsection{Energy balance and boundary layers}\label{eq:sec3b}

In a statistical steady-state, the external force performs work at a rate given by $\varepsilon = \left\langle \frac{1}{L^2}\int d^2 r\, \bm f \cdot \bm v \right\rangle$, which is balanced by the viscous dissipation in the system, expressed as:
\begin{equation}
    \varepsilon = 2 \nu \langle Z \rangle,
\end{equation}
where the angle brackets denote time averaging, and $Z = \frac{1}{L^2}\int d^2 \bm r\, \omega^2/2$ is the enstrophy. Both parts of this equality can be computed independently using DNS data, and they are indeed equal to each other with an accuracy of a few percent, see Table~\ref{tab:1}. The value of the pumping power ($\varepsilon = 5.66 \cdot 10^{-3}$ for DNS run $A$ and $\varepsilon = 4.59 \cdot 10^{-3}$ for DNS run $B$) also allows us to estimate the scale of enstrophy dissipation $l_d \sim \nu^{1/2} (\varepsilon k_f^2)^{-1/6}$ in the direct cascade. In both cases, this dissipation scale is found to be an order of magnitude larger than the grid size $L/N$ with $N=4096$.

\begin{figure}[t]
    \centering
    \includegraphics[width=0.8\linewidth]{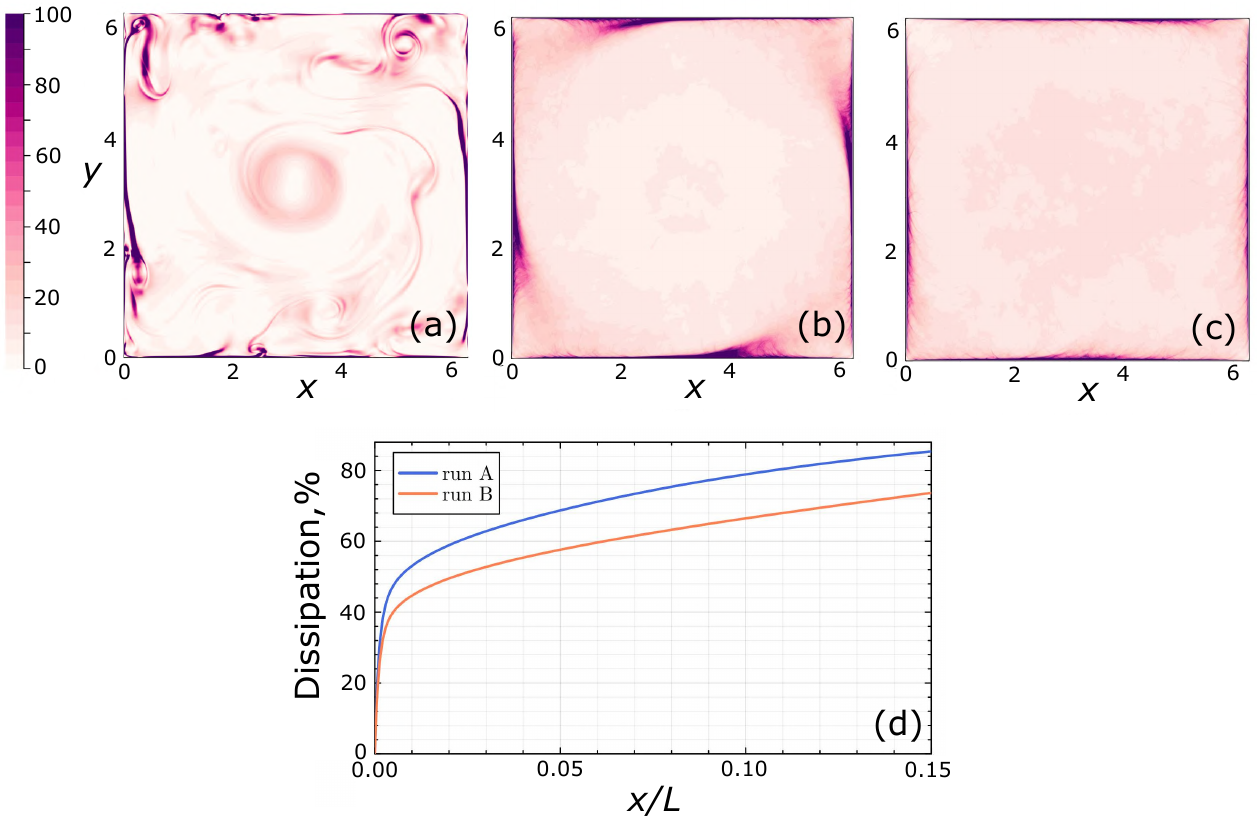}
    \caption{(a) The spatial density of viscous energy dissipation at a specific moment in time. (b,c) Time-averaged dissipation for DNS runs $A$ and $B$, respectively. (d) Cumulative dissipation fraction vs. normalized boundary layer width $x/L$.}
    \label{fig:3}
\end{figure}

The spatial density of viscous energy dissipation is given by $\nu (\partial_i v_k + \partial_k v_i)^2/(2L^2) $, and Fig.~\ref{fig:3}a shows the distribution of this quantity in space at a specific moment in time. Dissipation primarily occurs in narrow filaments near the boundaries, which then detach from the walls and extend into the fluid bulk, winding around the vortex structures. At the same time, dissipation is suppressed at the centers of the vortices, where the fluid undergoes solid-body rotation. Figures~\ref{fig:3}b and~\ref{fig:3}c show the time-averaged density of energy dissipation for DNS runs A and B, respectively. Note the asymmetry in Fig.~\ref{fig:3}b due to the presence of a coherent vortex in the center of the domain. Time averaging highlights that the dominant energy dissipation takes place near the edges of the system. Figure~\ref{fig:3}d quantifies this by showing the fraction of total energy dissipated within a boundary layer of width $x$, plotted as a function of $x/L$.

\begin{figure}[b]
    \centering
    \includegraphics[width=0.6\linewidth]{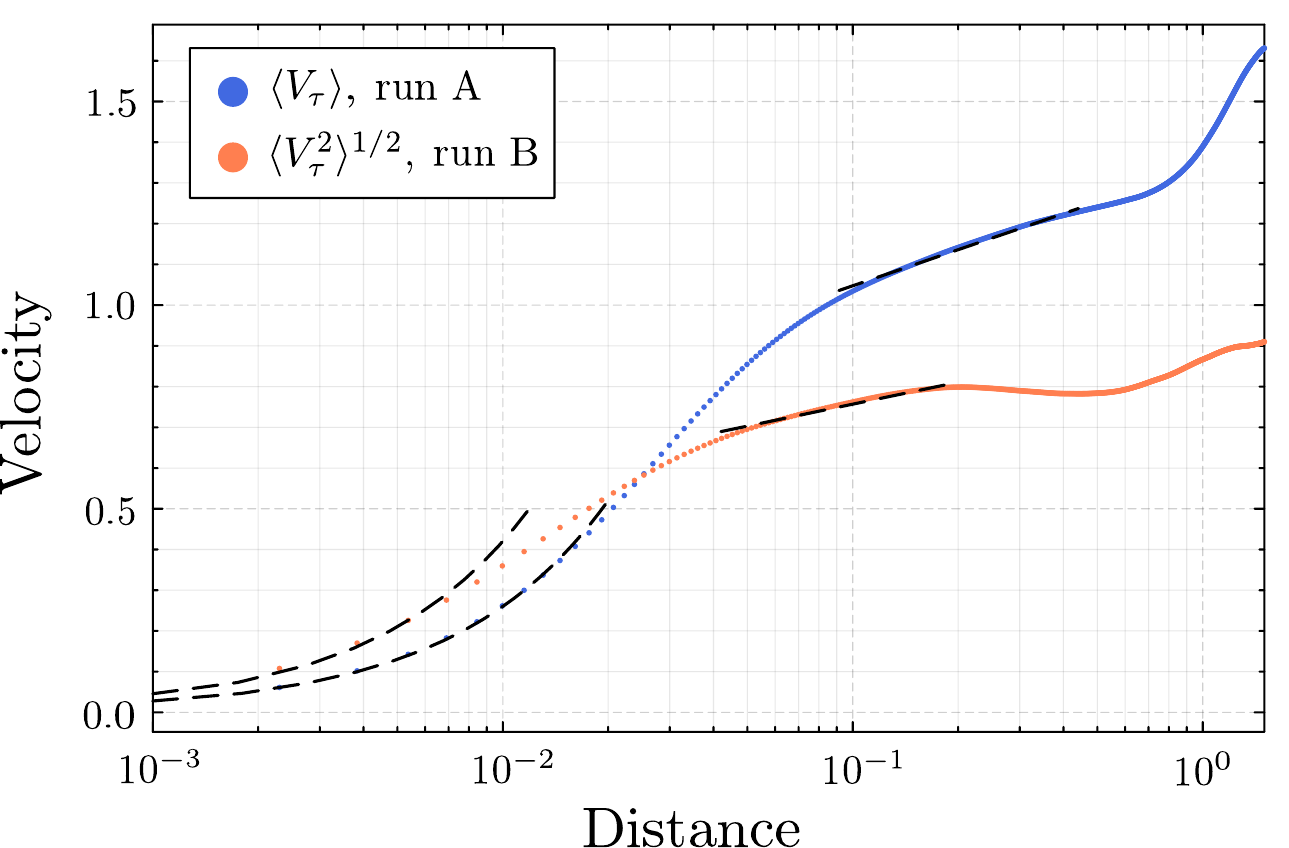}
    \caption{Velocity profiles in the boundary layer. Dashed lines indicate the linear and logarithmic regions. Note the semi-logarithmic scale.}
    \label{fig:4}
\end{figure}

Next, we examine the velocity profile in the boundary layer close to the wall, focusing on the cross-section passing through the centers of the sidewalls. For DNS run $A$, the average velocity along the wall $\langle V_{\tau} \rangle$ is non-zero and its direction is determined by the rotation of the coherent vortex at the center of the system. In contrast, for DNS run $B$, the average velocity is expected to be suppressed due to the reversals of the central vortex, and we focus on the root-mean-square profile of tangential velocity $\langle V_{\tau}^2 \rangle^{1/2}$. Figure~\ref{fig:4} shows how these quantities vary with the distance to the center of the domain wall, plotted on a semi-logarithmic scale. To increase statistics, we average the data near different walls. Close to the boundary, the fluid viscosity dominates and the velocity profiles exhibit a linear dependence on the distance from the wall --- this region can be identified as the linear viscous sublayer. Beyond this zone, the velocity profiles transition to a logarithmic dependence. However, the range of this logarithmic behavior is relatively limited, which can be attributed to the moderate Reynolds numbers in the simulations. These findings align with both experimental measurements of velocity profiles in soap films~\cite{samanta2014scaling} and earlier numerical studies using coarser grids~\cite{doludenko2021coherent}. It is worth noting that in our simulations, the grid resolution is fine enough to properly capture the viscous boundary sublayer.

\begin{figure}[t]
    \centering
    \includegraphics[width=0.8\linewidth]{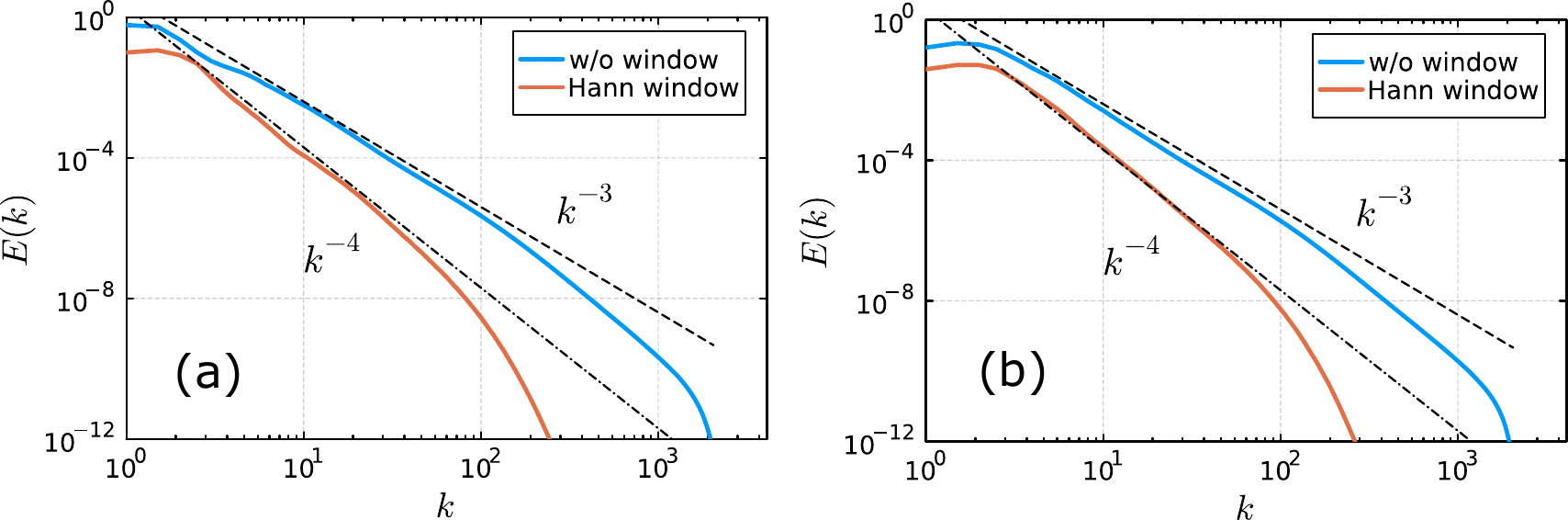}
    \caption{The energy spectra for DNS runs $A$ (a) and $B$ (b), both with and without the Hann window function.}
    \label{fig:5}
\end{figure}

The properties of turbulent flow are commonly described using the energy spectrum, as shown in Figs.~\ref{fig:5}a and \ref{fig:5}b for DNS runs $A$ and $B$, respectively. It is important to note that the analyzed system is spatially inhomogeneous due to the formation of boundary layers near the walls. As a result, the spectrum’s slope is significantly affected when the Hann window function is applied. Without windowing, the energy spectrum exhibits a slope consistent with Kraichnan’s prediction $E(k) \propto k^{-3}$. However, applying the Hann window, which effectively filters out the velocity field near the boundaries, results in a steeper spectral slope of approximately $E(k) \propto k^{-4}$. The results obtained are in qualitative agreement with the analysis of the second-order structure function of the vorticity presented in Ref.~\onlinecite{kramer2011structure}, which similarly indicates a steeper dependence $E(k) \propto k^{-4.5}$ within the central region of the domain. The enhanced spectral steepness may be associated with the presence of large-scale vortices near the cell center.

\subsection{Scaling behavior}\label{eq:sec3c}

In the previous section, we showed that the significant part of energy dissipation takes place near the no-slip boundaries of the system. This alters the scaling relations compared to the case where dissipation is concentrated in the fluid bulk, as is typical for homogeneous systems with periodic boundaries. The homogeneous case was previously examined theoretically in Ref.~\onlinecite{doludenko2021coherent}. Here, we consider how the corresponding estimates must be revised for systems with no-slip walls in the limit of large $Re$.


The estimate for the energy input rate remains the same and it can be expressed as $\varepsilon \sim f_0^2 \tau$, where $f_0$ denotes the forcing amplitude and $\tau$ is the characteristic timescale over which the flow induced by the forcing remains correlated with the forcing itself~\cite{tsang2009forced, doludenko2021coherent}. Despite the external force being steady in time, the correlation time $\tau$ is finite due to the temporal evolution of the flow. This timescale can be estimated as $\tau \sim (k_f U)^{-1}$, where $U$ represents the velocity of large-scale fluctuations and $1/k_f$ is the forcing scale.

In the statistically steady-state, the rate of energy input must balance the rate of energy dissipation, which predominantly occurs in thin boundary layers of thickness $\delta$ adjacent to the walls in the limit of large $Re$. This balance can be expressed as $\varepsilon L^2 \sim \nu (U/\delta)^2 \delta L$. To estimate the boundary layer thickness $\delta$, we compare the magnitudes of the nonlinear and viscous terms in the Navier–Stokes equation, yielding $\delta \sim \sqrt{\nu L/U}$. This estimate has been validated in previous numerical studies examining vortex dipole collisions with no-slip walls~\cite{clercx2017dissipation} and turbulent channel flows~\cite{falkovich2018turbulence}. Moreover, for the friction factor one can obtain $fr = \tau/(\rho U^2) \sim Re^{-1/2}$ in agreement with Ref.~\onlinecite{tran2010macroscopic}, where $\tau \sim \rho \nu U/\delta$ is the wall shear stress. Consequently, we obtain the following scaling relations:
\begin{eqnarray}
\label{eq:e_scaling}
&\varepsilon \sim f_0^{10/7} \nu^{1/7} k_f^{-5/7} L^{-3/7},&\\
\label{eq:U_scaling}
&U \sim f_0^{4/7} \nu^{-1/7} k_f^{-2/7} L^{3/7},&\\
\label{eq:d_scaling}
&\delta \sim f_0^{-2/7} \nu^{4/7} k_f^{1/7} L^{2/7}.&
\end{eqnarray}

\begin{figure}[t]
    \centering
    \includegraphics[width=0.8\linewidth]{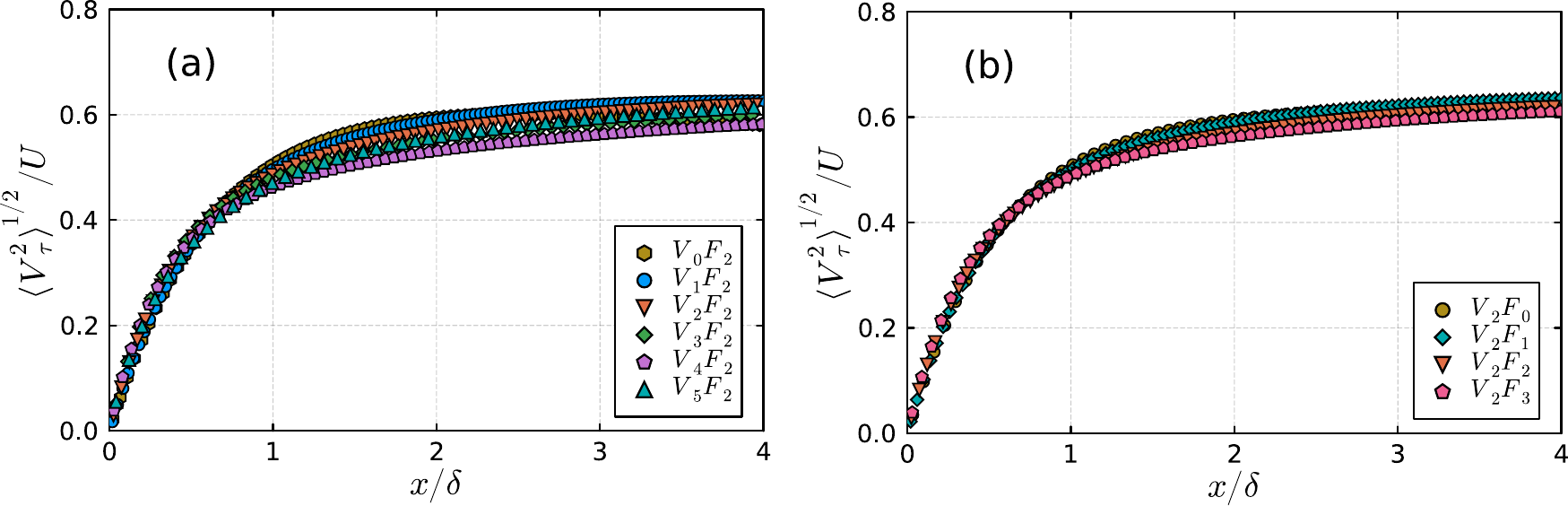}
    \caption{Dimensionless root-mean-square profile of tangential velocity within the boundary layer. The DNS results for runs $V_0 F_2$-$V_5 F_2$ (a) and $V_2 F_0$-$V_2 F_3$ (b) collapse onto a single universal curve, validating the scaling laws (\ref{eq:U_scaling}) and (\ref{eq:d_scaling}) used for the non-dimensionalization.}
    \label{fig:6}
\end{figure}

\begin{figure}[b]
    \centering
    \includegraphics[width=0.8\linewidth]{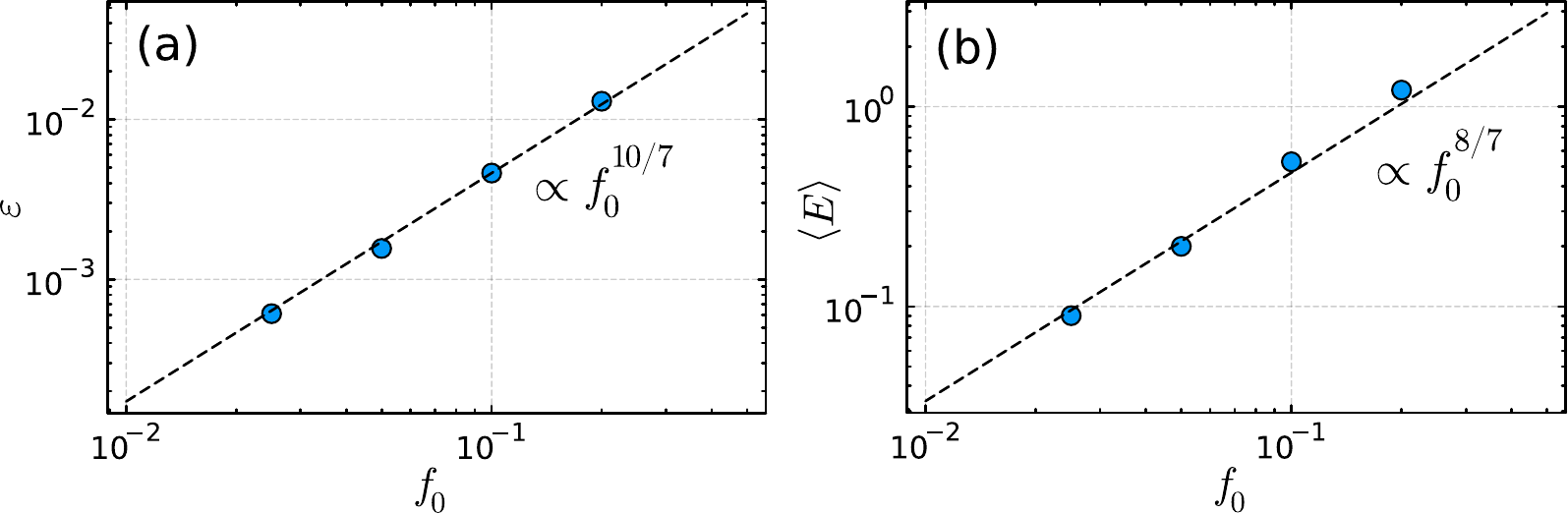}
    \caption{The dependence of (a) the pumping power $\varepsilon$ and (b) the mean energy $\langle E \rangle$ on the forcing amplitude $f_0$ for DNS runs $V_2 F_0$-$V_2 F_3$. The dashed lines represent the scaling according to the power law based on the expressions (\ref{eq:e_scaling}) and (\ref{eq:U_scaling}).}
    \label{fig:7}
\end{figure}

To support our findings, we performed DNS runs $V_0 F_2$-$V_5 F_2$ and $V_2 F_0$-$V_2 F_3$, in which the fluid viscosity $\nu$ and the forcing amplitude $f_0$ were varied by $32$ and $8$ times, respectively, see Table~\ref{tab:1} for details. We apply the same forcing (\ref{eq:fb}) as in DNS run $B$, which results in zero torque. For each case, we computed the root-mean-square profiles $\langle V_{\tau}^2 \rangle^{1/2}$ of tangential velocity near the boundary as a function of the distance to the wall in the cross-section passing through the centers of the sidewalls, and we average the data near different walls to increase statistics. These profiles were then presented in dimensionless form using the scaling expressions (\ref{eq:U_scaling}) and (\ref{eq:d_scaling}). The results are shown in Fig.~\ref{fig:6}, and the applied non-dimensionalization enables the data from different DNS runs to collapse onto a single master curve. Surprisingly, the curves collapse well even at fairly low $Re$, when a comparable fraction of the energy dissipation occurs in the cell volume away from the walls. Note that in all simulations the grid resolution in fine enough to resolve the boundary layer thickness $\delta$ and the scale of enstrophy dissipation $l_d \sim \nu^{1/2} (\varepsilon k_f^2)^{-1/6}$.

\begin{figure}[t]
    \centering
    \includegraphics[width=0.8\linewidth]{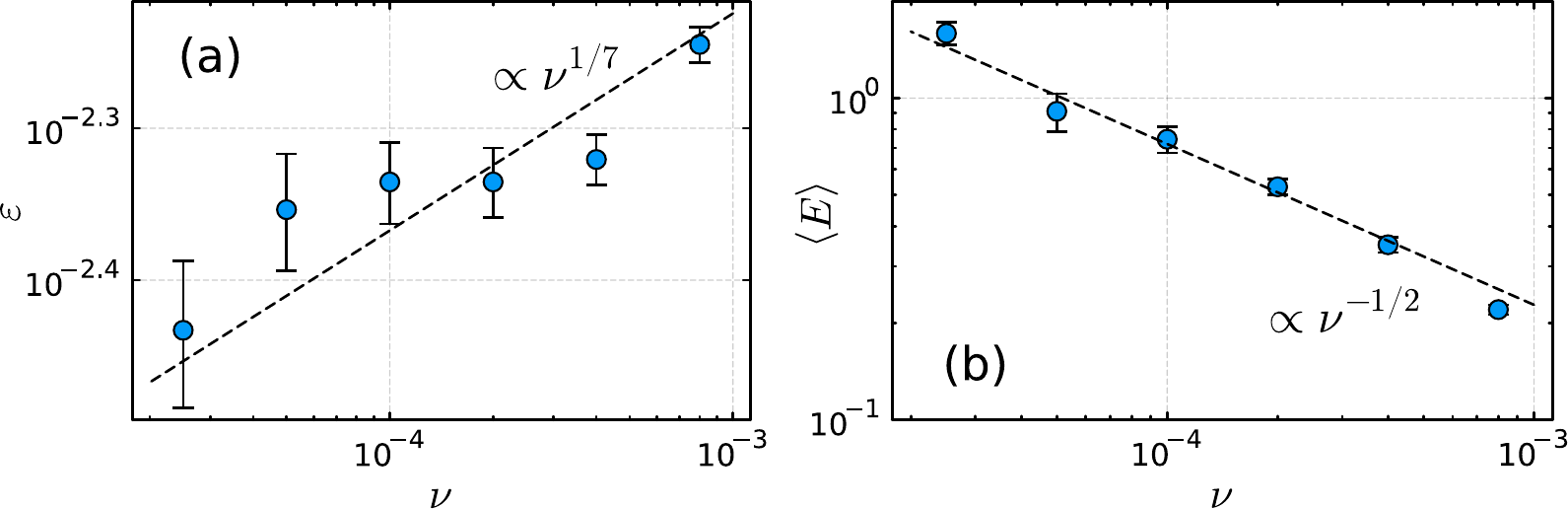}
    \caption{The dependence of (a) the pumping power $\varepsilon$ and (b) the mean energy $\langle E \rangle$ on the fluid viscosity $\nu$ for DNS runs $V_0 F_2$-$V_5 F_2$. The error bars are based on the central limit theorem and correspond to $\pm 3 \sigma / \sqrt{T / \max(\Delta t, \tau)}$, where $\sigma$ is the standard deviation, $T$ is the total duration of data collection, $\Delta t$ is the data sampling interval, and $\tau$ is the exponential decay time of the pair correlation function.}
    \label{fig:8}
\end{figure}

Figure~\ref{fig:7} shows the dependence of the pumping power $\varepsilon$ and the mean energy $\langle E \rangle$ on the forcing amplitude $f_0$ in logarithmic scale for DNS runs $V_2 F_0$-$V_2 F_3$. The results are consistent with the scaling relations (\ref{eq:e_scaling}) and (\ref{eq:U_scaling}). However, we should note that the predicted dependencies are close to their analogs for homogeneous system~\cite{doludenko2021coherent} $\varepsilon \propto f_0^{4/3}$ and $\langle E \rangle \propto f_0^{4/3}$. The direct verification of the scaling (\ref{eq:e_scaling}) and (\ref{eq:U_scaling}) with respect to fluid viscosity $\nu$ is more challenging due to the weak sensitivity of these quantities to its variations. The results from DNS runs $V_0 F_2$-$V_5 F_2$ are shown in Fig.~\ref{fig:8}. The observed dependencies are indeed flatter than the corresponding predictions $\varepsilon \propto \nu^{1/3}$ and $\langle E \rangle \propto \nu^{-2/3}$ for homogeneous systems~\cite{doludenko2021coherent}. The pumping power demonstrates the behavior in reasonable agreement with our prediction $\varepsilon \propto \nu^{1/7}$, while the mean energy follows an approximate scaling of $\langle E \rangle \propto \nu^{-1/2}$. The observed exponent lies between our theoretical estimate $-2/7$ and the classical homogeneous case $-2/3$, suggesting that the simulations have not yet reached sufficiently high Reynolds numbers to fully capture the transition to the new scaling regime. Note also that the energy dissipation within the boundary layer of width $\delta$ accounts only for nearly $40 \%$ of the total dissipation at $\nu=2\cdot 10^{-4}$, as shown in Fig.~\ref{fig:3}d. Overall, the proposed scaling calls for more careful validation, which requires additional DNS at higher Reynolds numbers.

\subsection{Torque to no-torque transition}\label{eq:sec3d}

In this section, we explore how the system's behavior changes as the external force torque varies smoothly. Our objective is to perform long-term DNS to capture multiple reversals in the rotation direction of the central vortex. The external force is expressed as:
\begin{equation}
    \bm f_X = [f_0 \sin(k_f y), f_0 \sin(k_f x + \phi)],
\end{equation}
and by adjusting the phase $\phi \in [0, \pi]$, we control the torque $M$ relative to the cell center $\bm r_O$:
\begin{equation}
    M = \frac{1}{L^2}\int d^2 \bm r\, [(\bm r - \bm r_O)\times \bm f_X]_z =\dfrac{2 f_0}{k_f} \sin^2 (\phi/2).
\end{equation}
We set the forcing amplitude $f_0 = 0.1$, the pumping wavenumber $k_f = 5$, and the fluid viscosity $\nu = 2 \cdot 10^{-4}$. Grid convergence tests indicate that the minimum resolution required is $1024^2$, which allows us to observe the system's steady-state dynamics over a time period of $T = 10^5$. We output data at every $\Delta t =20$ that is small compared to the typical lifetime of the central vortex. The parameters of the performed DNS runs $X_0$-$X_{12}$ are summarized in Table~\ref{tab:1}. Note that the external force exciting the flow has a different spatial structure for DNS runs $X_0$ (with $\phi=0$) and $B$, so the statistical characteristics for these calculations do not coincide, but one can compare DNS runs $X_{12}$ (with $\phi=\pi$) and $A$ to ensure that the grid resolution is sufficient.

\begin{figure}[t]
    \centering
    \includegraphics[width=0.7\linewidth]{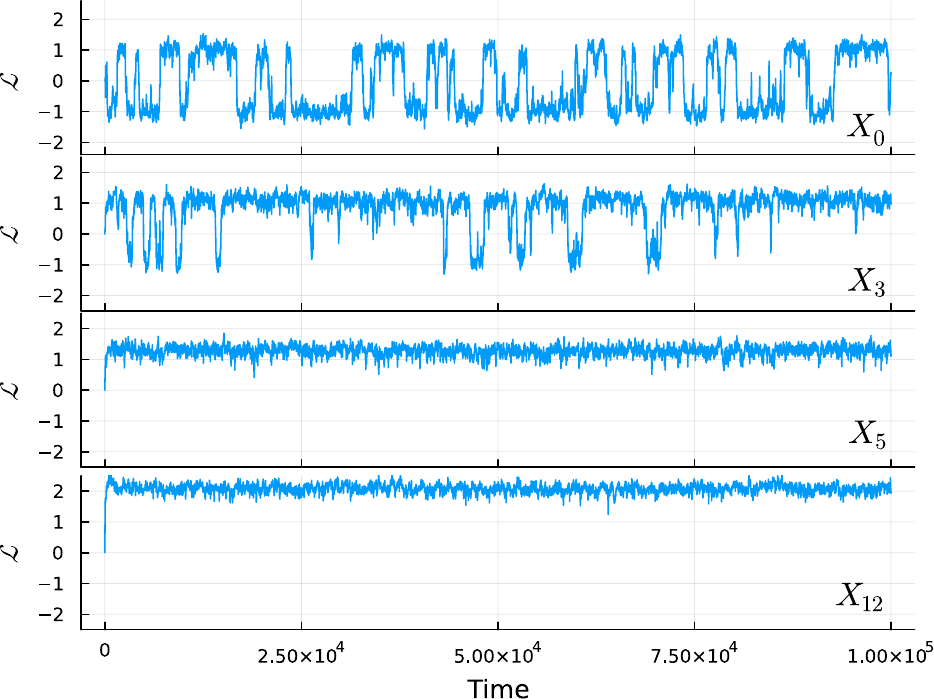}
    \caption{Time evolution of angular momentum $\mathcal{L}$ for DNS runs $X_0, \; X_3, \; X_5, \; X_{12}$.  }
    \label{fig:9}
\end{figure}

Figure~\ref{fig:9} illustrates the time evolution of angular momentum $\mathcal{L}$ across several DNS runs. When the external force torque is relatively large, the angular momentum remains consistently positive, indicating the presence of a stable central vortex (DNS runs $X_5$--$X_{12}$). At lower torques, the angular momentum can randomly change sign, reflecting spontaneous reversals in the vortex's rotation (DNS runs $X_0$--$X_4$). Despite these reversals, the system predominantly resides in states where the angular momentum shares the sign of the applied torque. In the absence of torque, no preferred direction of rotation emerges; the average angular momentum is close to zero, and central vortices of either sign appear with equal probability.

\begin{figure}[b]
    \centering
    \includegraphics[width=0.8\linewidth]{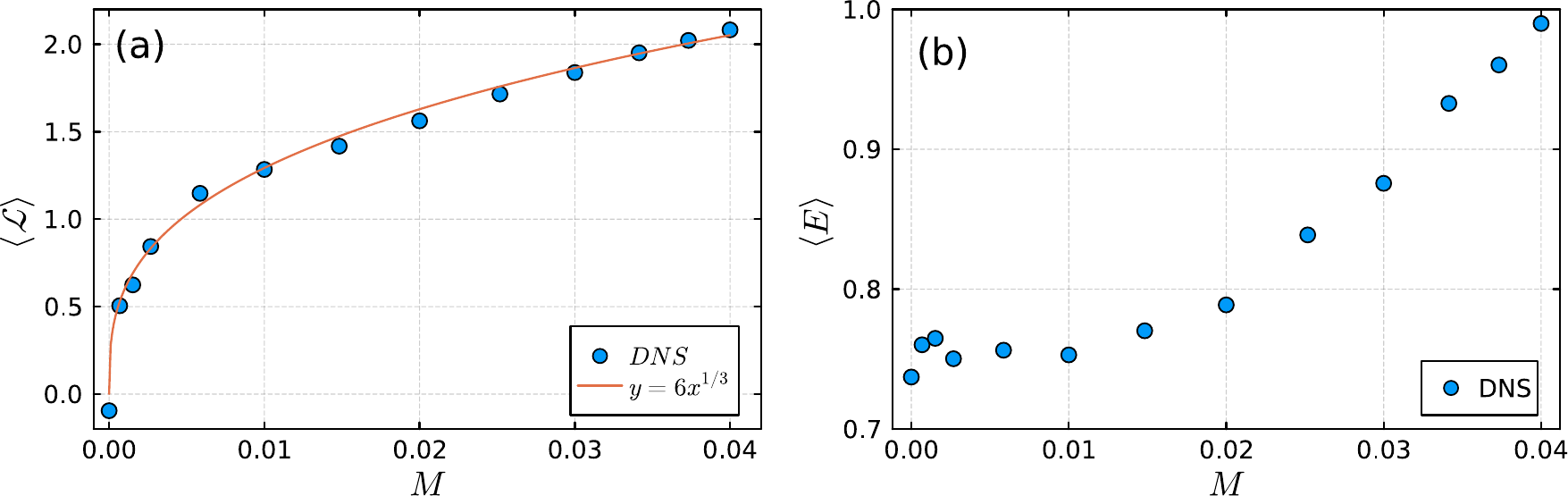}
    \caption{Dependencies of (a) the mean angular momentum $\langle \mathcal{L} \rangle$ and (b) the mean energy $\langle E \rangle$ on the applied torque $M$. The orange line shows an approximate scaling $\langle \mathcal{L} \rangle \propto M^{1/3}$.}
    \label{fig:10}
\end{figure}

Next, we examine how the mean angular momentum $\langle \mathcal{L} \rangle$ and the mean energy $\langle E \rangle$ in the statistical steady-state depend on the applied torque $M$, see Fig.~\ref{fig:10}. Both dependencies appear to be smooth. The data for the mean angular momentum follows an approximate relationship $\langle \mathcal{L} \rangle \propto M^{1/3}$. For relatively large torques, variations in the mean angular momentum correspond to changes in its typical value and are accompanied by a change in the mean energy of the flow. For relatively small torques, the flow energy and the typical amplitude of angular momentum remain nearly constant (see Fig.~\ref{fig:10}b and Fig.~\ref{fig:9}, respectively) and changes in the mean angular momentum arise primarily from spontaneous reversals of the central vortex's rotation.

An intriguing question is how the transition between the described regimes occurs -- whether it is smooth or exhibits a threshold behavior? In the first case, angular momentum could spontaneously change sign at any external torque, but such events are not observed in simulations with large torques. This is because the simulations are time-limited, and the probability of a sign reversal is extremely low. In the second case, sign reversals are prohibited when the applied torque $M$ exceeds a certain threshold value $M^*$. The smooth dependencies shown in Fig.~\ref{fig:10} seem to favor the first scenario, though they are not definitive enough to provide a conclusive answer. A theoretical explanation for the observed behavior, as well as the dependence of the threshold value $M^*$ (if one exists) on the system parameters, remains an open question for future research. 

\section{Conclusion}

In this work, we have investigated the effects of no-slip boundaries and external force torque on two-dimensional turbulence within a square domain using high-resolution direct numerical simulations. Our findings highlight the significant role played by both boundary conditions and the torque $M$ of the external forcing in shaping the large-scale dynamics of the system. When $M$ is relatively large, a stable long-lived vortex emerges at the center of the domain, sustaining a persistent angular momentum. At lower torques, the angular momentum exhibits random reversals in sign, driven by spontaneous switches in the vortex's circulation direction. Despite these reversals, the angular momentum tends to align with the direction of the applied torque. In the absence of torque, no rotational preference is established, and vortices of both signs appear with equal probability at the center. 
Remarkably, the time-averaged angular momentum follows a smooth scaling law $\langle \mathcal{L} \rangle \propto M^{1/3}$, without a signature of different regimes.
These findings reveal a rich dynamical response of the system to changes in forcing asymmetry and motivate further exploration of the transition between regimes.



The significant part of energy pumped into the system is found to be dissipated near the boundaries, necessitating a revision of traditional scaling laws for homogeneous systems. We derived new scaling relations for the energy input rate (\ref{eq:e_scaling}), characteristic velocity (\ref{eq:U_scaling}), and boundary layer thickness (\ref{eq:d_scaling}) that account for the influence of the no-slip walls. These relations were validated through DNS with varying fluid viscosities and forcing amplitudes, demonstrating a collapse of tangential velocity profiles in the boundary layers close to the walls onto a universal curve when appropriately non-dimensionalized. The velocity profiles across boundary layers exhibit a well-defined structure: a linear viscous sublayer adjacent to the wall, followed by a region of logarithmic dependence.

Although this study focused on the case of stationary forcing, the derived scaling relations can be readily extended to the scenario of stochastic forcing that is short-time correlated. In such cases, the energy input rate $\varepsilon$ becomes a prescribed external parameter, which can be directly controlled in numerical simulations. Using the previously established estimate for the boundary layer thickness $\delta \sim \sqrt{\nu L/U}$ and equating the viscous energy dissipation to the input power $\varepsilon L^2 \sim \nu (U/\delta)^2 L \delta$, we obtain the following estimates:
\begin{eqnarray}
    &U \sim \varepsilon^{2/5} L^{3/5} \nu^{-1/5},&\\
    &\delta \sim \varepsilon^{-1/5} L^{1/5} \nu^{3/5}.&
\end{eqnarray}
The consideration above implies that the characteristic velocity $U$ exceeds the typical velocity scale $(\varepsilon L)^{1/3}$ of Kolmogorov fluctuations at the system size $L$. This means that the inverse energy cascade extends up to the system scale $L$, resulting in energy accumulation at that scale. The scenario corresponds to a large Reynolds number for Kolmogorov fluctuations, expressed as $\varepsilon^{1/3} L^{4/3} / \nu \gg 1$, which always holds if the Reynolds number for the pumping scale is large. Thus, friction at the boundaries is not sufficient to halt the inverse energy cascade before it reaches the system size. Verification of this qualitative picture in simulations with stochastic forcing and no-slip boundary conditions requires further studies. The results presented in this work enhance our understanding of how boundary layers and forcing asymmetry control large-scale structure formation in 2D turbulence and offer a foundation for future work in confined turbulent systems.

\acknowledgments

We thank Vladimir Lebedev, Igor Kolokolov, and Sergey Vergeles for helpful discussions. The work was supported by the Russian Science Foundation (Project No. 23-72-30006). The authors are grateful to the Landau Institute for providing computing resources.

\section*{Data AVAILABILITY}
The data that support the findings of this study are available from the corresponding author upon reasonable request.

\bibliography{biblio}

\end{document}